\documentclass[sigconf]{acmart}

\AtBeginDocument{%
  }

\copyrightyear{2025}
\acmYear{2025}
\setcopyright{cc}
\setcctype{by}
\acmConference[IUI '25]{30th International Conference on Intelligent User Interfaces}{March 24--27, 2025}{Cagliari, Italy}
\acmBooktitle{30th International Conference on Intelligent User Interfaces (IUI '25), March 24--27, 2025, Cagliari, Italy}\acmDOI{10.1145/3708359.3712092}
\acmISBN{979-8-4007-1306-4/25/03}

\usepackage{url}
\usepackage{pifont} 
\usepackage{booktabs}
\usepackage{multirow}
\usepackage{tabularx}
\usepackage{xcolor}
\usepackage{xspace}
\usepackage{listings}
\usepackage{xstring}
\usepackage[htt]{hyphenat}
\usepackage{subcaption}
\usepackage{makecell}
\usepackage{colortbl}
\usepackage{enumitem}
\usepackage{pifont}
\usepackage{tikz}
\usepackage{enumitem}
\usepackage{framed}
\usepackage{soul}

\usepackage{fontawesome}
\usepackage{pgfplots}
\usepackage{pgfplotstable}
\usepackage{multirow}
\usepackage{multicol}
\usepackage{changepage}

\usepackage{acmart-taps}
\usepackage{framed}

\newcommand{\toolname}[0]{\textsc{Orbit}}
\newcommand\circleone{\ding{192}}
\newcommand\circletwo{\ding{193}}
\newcommand\circlethree{\ding{194}}
\newcommand\circlefour{\ding{195}}
\newcommand\circlefive{\ding{196}}
\newcommand\circlesix{\ding{197}}
\newcommand\circleseven{\ding{198}}
\newcommand\circleeight{\ding{199}}

\newcommand{\new}[1]{{#1}}

\definecolor{backg}{RGB}{225,236,244}
\definecolor{tagtxt}{RGB}{88,115,159}
\definecolor{darkgrey}{HTML}{A9A9A9}

\newcommand\objective[1]{%
    \textcolor{black}{{\texttt{#1}}}\xspace
}

\definecolor{metricbg}{RGB}{200,230,201}
\definecolor{metrictxt}{RGB}{85, 107, 47}

\newcommand\metric[1]{%
    \textcolor{black}{\texttt{#1}}\xspace
}

\definecolor{deeporange}{HTML}{e6550d}
\definecolor{lightorange}{HTML}{fdae68}
\definecolor{deepgreen}{HTML}{74c476}
\definecolor{lightgreen}{HTML}{c7e9c0}
\definecolor{deepblue}{HTML}{6baed6}
\definecolor{lightblue}{HTML}{c6dbef}

\definecolor{orange}{HTML}{ff7f0e}
\definecolor{blue}{HTML}{1f77b4}

\newcommand{\bigstep}{\raisebox{3pt}{\fcolorbox{white}{deeporange}{}}}
\newcommand{\smallstep}{\raisebox{3pt}{\fcolorbox{white}{lightorange}{}}}
\newcommand{\addmetric}{\raisebox{3pt}{\fcolorbox{white}{deepgreen}{}}}
\newcommand{\stdmetric}{\raisebox{3pt}{\fcolorbox{white}{lightgreen}{}}}
\newcommand{\addexm}{\raisebox{3pt}{\fcolorbox{white}{deepblue}{}}}
\newcommand{\stdexm}{\raisebox{3pt}{\fcolorbox{white}{lightblue}{}}}

\newcommand{\treatment}{\raisebox{2pt}{\fcolorbox{white}{orange}{}}}
\newcommand{\control}{\raisebox{2pt}{\fcolorbox{white}{blue}{}}}

\definecolor{PAblue}{RGB}{0,122,204}%
\definecolor{PAlightblue}{RGB}{235, 247, 255}%

\newcommand{\quo}[1]{ 
	\vspace{-0.1cm}
	\def\FrameCommand{%
		\hspace{8pt}%
		{\color{PAblue}\vrule width 2pt}%
		{\color{white}\vrule width 2pt}%
		\colorbox{white}
	}%
	\MakeFramed{\advance\hsize-\width\FrameRestore}%
	\noindent\hspace{-4.55pt}%
	\begin{adjustwidth}{}{0pt}
		{#1}
		\vspace{-3pt}
	\end{adjustwidth}\endMakeFramed%
}

\definecolor{green3}{RGB}{66,179,130}
\definecolor{green2}{RGB}{121,205,169}
\definecolor{green1}{RGB}{196,233,217}

\definecolor{red1}{RGB}{251,219,220}
\definecolor{red2}{RGB}{244,164,166}
\definecolor{red3}{RGB}{236,91,96}

\definecolor{blue1}{RGB}{101,173,246}
\definecolor{blue2}{RGB}{12,112,212}

\definecolor{orange1}{RGB}{250,181,97}
\definecolor{orange2}{RGB}{245,138,7}

\definecolor{purple1}{RGB}{195,176,232}
\definecolor{purple2}{RGB}{146,112,212}

\definecolor{pink1}{RGB}{236,152,223}
\definecolor{pink2}{RGB}{227,100,208}

\definecolor{gray1}{RGB}{220,220,220}

\newcommand{\participantQuote}[2]{\textit{``#1'' (P#2)}}

\usepackage{iui25-22}

\begin{document}

\title{Orbit: A Framework for Designing and Evaluating Multi-objective Rankers}

\author{Chenyang Yang}
\authornote{Work done at Amazon.}
\affiliation{%
  \institution{Carnegie Mellon University}\country{}
}

\author{Tesi Xiao}
\authornote{Equal contribution.}
\affiliation{%
  \institution{Amazon}\country{}
}

\author{Michael Shavlovsky}
\authornotemark[2]
\affiliation{%
  \institution{Amazon}\country{}
}

\author{Christian K\"astner}
\affiliation{%
  \institution{Carnegie Mellon University}\country{}
}

\author{Tongshuang Wu}
\affiliation{%
  \institution{Carnegie Mellon University}\country{}
}

\renewcommand{\shortauthors}{Yang et al.}

\begin{abstract}
Machine learning in production needs to balance multiple objectives:
This is particularly evident in ranking or recommendation models, where conflicting objectives such as user engagement, satisfaction, diversity, and novelty must be considered at the same time. 
However, designing multi-objective rankers is inherently a dynamic wicked problem -- there is no single optimal solution, and the needs evolve over time. 
Effective design requires collaboration between cross-functional teams and careful analysis of a wide range of information.
In this work, we introduce Orbit, a conceptual framework for Objective-centric Ranker Building and Iteration.
The framework places objectives at the center of the design process, to serve as boundary objects for communication and guide practitioners for design and evaluation.
We implement Orbit as an interactive system, which enables stakeholders to interact with objective space directly and supports real-time exploration and evaluation of design trade-offs. 
We evaluate Orbit through a user study involving twelve industry practitioners, showing that it supports efficient design space exploration, leads to more informed decision-making, and enhances awareness of the inherent trade-offs of multiple objectives. 
Orbit (1) opens up new opportunities of an objective-centric design process for any multi-objective ML models, as well as (2) sheds light on future designs that push practitioners to go beyond a narrow metric-centric or example-centric mindset.
\end{abstract}



\maketitle

\section{Introduction}
Machine learning models are ubiquitous, yet it can be hard to do them right in production~\cite{NZLK:ICSE22}.
To train a model, it is customary to define appropriate training \textit{objectives}:
For a language model, the training objective is to predict next tokens, with every token serving as the training target for the context before.
For an image classification model, the training objective is to predict the correct label annotated. %
However, in many cases, a model has \textit{more than one objective} to train for, with the most prominent example being ranking or recommendation models~\cite{rodriguez2012multiple}.

Take video recommendation as an example:
There are many user behavioral signals that can be used as objectives,
from \objective{clicks} and \objective{watch time} to capture user engagement,
to \objective{likes} and \objective{ratings} to capture user satisfaction~\cite{zhao2019recommending}.
Beyond recommendation accuracy, objectives like \objective{diversity}, \objective{serendipity}, \objective{novelty}, and \objective{coverage}~\cite{kaminskas2016diversity} can also be incorporated into model training,
such as to avoid filter bubbles~\cite{pariser2011filter}, encourage user exploration~\cite{toms2000serendipitous}, and ultimately improve long-term user experience.
These different objectives can easily conflict with each other~\cite{zheng2022survey} and might be prioritized differently by different stakeholders~\cite{surer2018multistakeholder}.
End users might care more about \objective{likes} and \objective{ratings} to find enjoyable content, while content creators might care more about \objective{diversity} and \objective{novelty} for better video visibility.
Advertisers might prioritize \objective{click} to maximize ads visibility, while content creators would favor higher \objective{watch time} for their content.
With only one single ranking order that can be produced, there is always a decision to make on what items should go above others -- and there is no strict best order for that decision~\cite{rodriguez2012multiple}.

\textit{Designing multi-objective rankers is, therefore, inherently a wicked problem}~\cite{rittel1973dilemmas}: 
There is no definitive formulation, no stopping rules, and no single ``best'' solution.
There are always choices on what objectives to incorporate and how to trade off different objectives. 
Furthermore, this wicked problem of ranker design is \textit{dynamic} with changing stakeholder needs.
For example, Youtube's video recommenders evolved from considering only watch time~\cite{covington2016deep}, to multiple user behavioral signals capturing user satisfaction beyond engagement~\cite{zhao2019recommending}, and more recently to incorporating diversity as part of the objectives to mitigate echo chamber effect~\cite{wang2024diversifying}.
For a ranking system in production, there are constantly new observations and feedback from stakeholders that motivate the need for \textit{continuous (re-)design of multi-objectives rankers}. 

To thoroughly consider trade-offs for ranker design, practitioners need significant efforts both (a) analyzing and incorporating feedback from different stakeholders and (b) looking across various kinds of evidence.

\paragraph{Communication and collaboration}
First, as is typical with many other machine learning systems~\cite{NZLK:ICSE22}, designing multi-objective rankers is a \textit{cross-functional} efforts: 
We observed (cf. Section~\ref{sec:motivation-collab}) that, 
various stakeholders (e.g., product managers) frequently provide \textit{feedback} on where a ranker can improve, and technical stakeholders (e.g., ML engineers, scientists) will need to translate the feedback into appropriate updates to model objectives and re-design the model.
However, it can be challenging for different stakeholders to effectively communicate and collaborate:
Less technical stakeholders can struggle to provide actionable feedback, while technical stakeholders have to spend significant efforts to analyze their feedback and incorporate it into the ranker if plausible.

\paragraph{Design and evaluation}
Second, designing and evaluating multi-objective rankers is an endeavor involving careful analysis of a rich set of information: 
We observed (cf. Section~\ref{sec:motivation-eval}) that stakeholders need to track aggregated \textit{metrics} to understand overall trends of each objective, 
inspect concrete \textit{examples} to understand users' concrete experiences,
and also inspect data \textit{slices}~\cite{cabrera2023zeno} to analyze important subgroups and more nuanced phenomena.
Tracking all the information at the same time is challenging, and makes practitioners struggle to design appropriate rankers.

In this work, we propose a conceptual framework, \toolname{}, for \textit{Objective-centric Ranker Building and Iteration}.
The key idea is that \textbf{objectives should take the central role in the model design process, to guide communication, exploration, and evaluation}. 
We argue that objectives can act as the boundary object~\cite{boundary-object} between stakeholders,
to be interpreted colloquially and connected to stakeholder feedback and concrete examples,
and also to be defined precisely in mathematical terms for model training \new{(cf. Section~\ref{sec:definition})}.
For practitioners designing and evaluating multi-objective rankers, we argue that objectives can help navigate design space and forage information for evaluation, as they define where to explore, inform what to evaluate, and explicate the inherent trade-offs.

We implemented \toolname{} as an interactive system that affords interactive ranker design: 
Users can directly operate on the objective space, and observe how concrete examples and aggregated metrics change in real time,
allowing much more efficient and well-informed exploration of the design space.
\toolname{} also serves as a platform for less techncial stakeholders to better understand multi-objective rankers and potentially provide more constructive feedback in the design process.
To evaluate \toolname{}, we conducted a user study with twelve experienced industry practitioners.
Our evaluation shows that with \toolname{}, users can explore the design space more efficiently, make more informed decisions, and are more likely to communicate the inherent trade-offs to other stakeholders.

To summarize, our work makes the following contribution:
\begin{itemize}
    \item A perspective of multi-objective ranker design as a dynamic wicked problem and its associated challenges identified in practice, which enable new design approaches.
    \item An objective-centered conceptual framework for multi-obje\-ctive ranker design that provides a foundation for our and future system design.
    \item \toolname{}, an interactive system supporting interactive ranker design and evaluation.
    \item Insights from user studies that objective-centered design supports users to explore the design space more efficiently, make more informed decisions, and be more aware of the trade-offs, shedding light on designing similar systems for other multi-objective ML problems.
\end{itemize}

\section{Motivation}
\label{sec:motivation}

We embedded ourselves in a team (n=50) responsible for commercial product rankers over five months.
The team members have a wide range of different roles, from applied scientists, software engineers, product managers, to machine learning engineers.
The team regularly updates new models on a monthly basis. %
We conducted informal interviews with team members, studied their existing workflows, as well as analyzed internal documents on past cross-functional communication on the product rankers.
Through this process, we identified two key challenges for ranker design in their day-to-day activities, which we summarize below.

\subsection{Communication and Collaboration: Lack of a Shared Language}
\label{sec:motivation-collab}
We first found that designing and evaluating rankings is not a one-side effort from technical stakeholders.
Lots of less technical stakeholders were involved in the past history and wanted to provide feedback on the ranker to improve user experiences on different data slices and dimensions.
This can be particularly helpful, as they often bring in domain expertise and provide feedback grounded in concrete observations and customer experiences.
However, even though these stakeholders have the domain expertise and some insights, we found they can struggle to provide \textit{actionable} feedback that can be incorporated in the next model iteration.

Indeed, the process of understanding and incorporating feedback is often perceived to be frustrating and sometimes not constructive -- it typically takes weeks of communication efforts.
This is because less technical stakeholders only have a vague notion of what objectives the current model is trained for --
they communicate what they \textit{want}, without understanding what can be achieved, especially with the constraints of balancing multiple objectives.
Meanwhile, incorporating their feedback into model design requires deliberate thinking on how one can translate and express feedback in the objectives, which can take lots of effort for technical stakeholders.
The lack of a shared language slows down communication and collaboration, and hinders faster iteration over model design.

This, along with our other observations in Section~\ref{sec:motivation-eval}, motivates our first design goal:
\begin{enumerate}
\item[{G1.}] \textit{Objective-centric.} \textbf{Objectives should be the first class citizen.} 
The current design process suffers from the lack of a shared language and appropriate guidance.
The system should surface objectives as the main object for stakeholders to navigate through the design space, communicate their findings, and negotiate over trade-offs.
\end{enumerate}

\subsection{Design and Evaluation: Plethora of Information}
\label{sec:motivation-eval}
To design rankers and evaluate a design, practitioners need to forage various information, from metrics, examples, to slices:
They need to track aggregated \textit{metrics} to understand the overall trends of each objective, mostly in a designated dashboard.
However, as mentioned by one of the practitioners, \textit{``...(our metric) aggregated NDCG is sparse and not always reliable,''}
echoing existing concerns on common information retrieval evaluation metrics~\cite[e.g.,][]{ferrante2017ir, moffat2022batch, ferrante2020interval, ferrante2021towards}, and generally discussion on how aggregated metrics can hide lots of nuances in machine learning~\cite[e.g.,][]{ribeiro-etal-2020-beyond}.
Therefore, practitioners also heavily rely on inspecting concrete examples to confirm whether the aggregated metric improvements aligned with human expectations, which is supported by an internal platform to inspect and analyze individual examples.
Stakeholders also conduct more customized analyses on important subsets (known as data slices~\cite{cabrera2023zeno}), for which they need to switch to computational notebooks for their expressive power.

This three-fold metric-example-slice information foraging process is much more complicated compared to an idealized machine learning setup, 
where stakeholders exclusively focus on optimizing models towards a well-defined objective and measure progress through aggregated metrics. 
In one of our observation sessions, we found the practitioner started with analyzing examples in an explanability tool, but quickly jumped to notebooks for more detailed analysis, and switched to an example analysis platform once an example was found interesting.
Because it is mentally challenging to forage comprehensive information for model design,
stakeholders often choose to test design hypotheses highly selectively, and only explore a few alternatives per design hypothesis. 
This leaves a large design space mostly unexplored, and potentially many iteration opportunities missing.
Sometimes, they do not have time to conduct comprehensive evaluations, leaving their decision-making up to a few key metrics, and lots of nuances unexplored.
These observations motivate our second design goal:
\begin{enumerate}
    \item[{G2.}] \textit{Comprehensive-evals.} \textbf{Evaluations should be comprehensive and cover different types of information.}
Users need evaluation results to assist their model design iteration.
The evaluations should be comprehensive, with both qualitative (examples) and quantitative (metrics, slices) information, allowing users to quickly assess a design.
\end{enumerate}

As mentioned above, we also observed that practitioners heavily rely on computational notebooks for any more customized and detailed analysis.
Computational notebooks are particularly powerful for their expressiveness, supporting practitioners to experiment with complicated objectives and define appropriate metrics for evaluations.
Such customizability is essential to ranker design, motivating our final design goal:

\begin{enumerate}
\item[{G3.}] \textit{Customizability.} \textbf{Objectives and metrics should be easily customizable.}
Practitioners often want to define refined objectives with complex interactions and customize what to evaluate. 
The system should support them to customize their design and evaluation, such that they can experiment with different designs and perform in-depth analysis.

\end{enumerate}

\begin{figure}[t]
    \centering
    \includegraphics[width=0.8\linewidth]{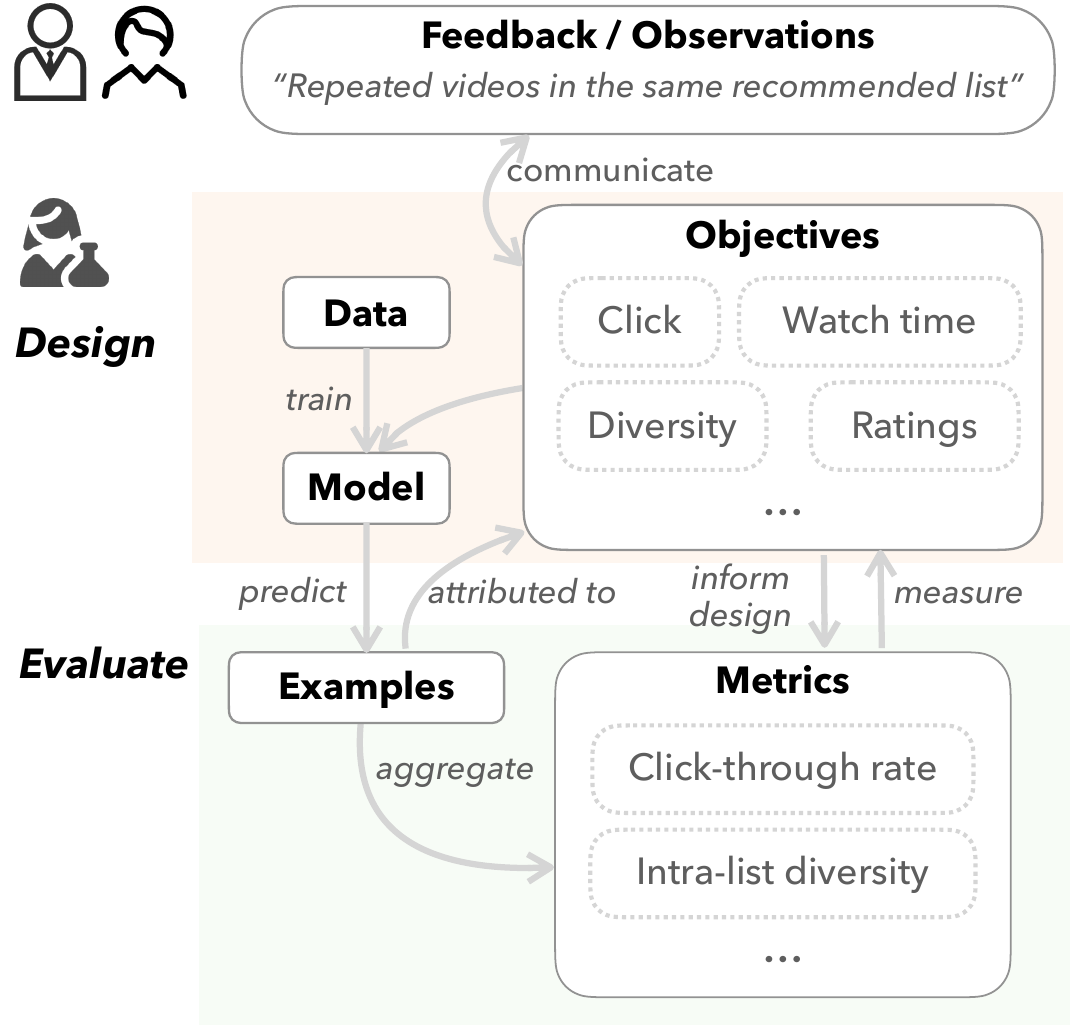}
    \caption{\toolname{}'s conceptual model. Objectives take the central role of ranker design: 
    They can be translated from feedback and concrete observations, serving as the bridge between different stakeholders.
    They can help practitioners explore different model designs.
    They can help conduct evaluation, by informing what metrics to design and track and providing attribution for concrete ranking results.
    }
    \label{fig:concept}
\end{figure}

\begin{figure*}[ht]
    \centering
    \includegraphics[width=\linewidth]{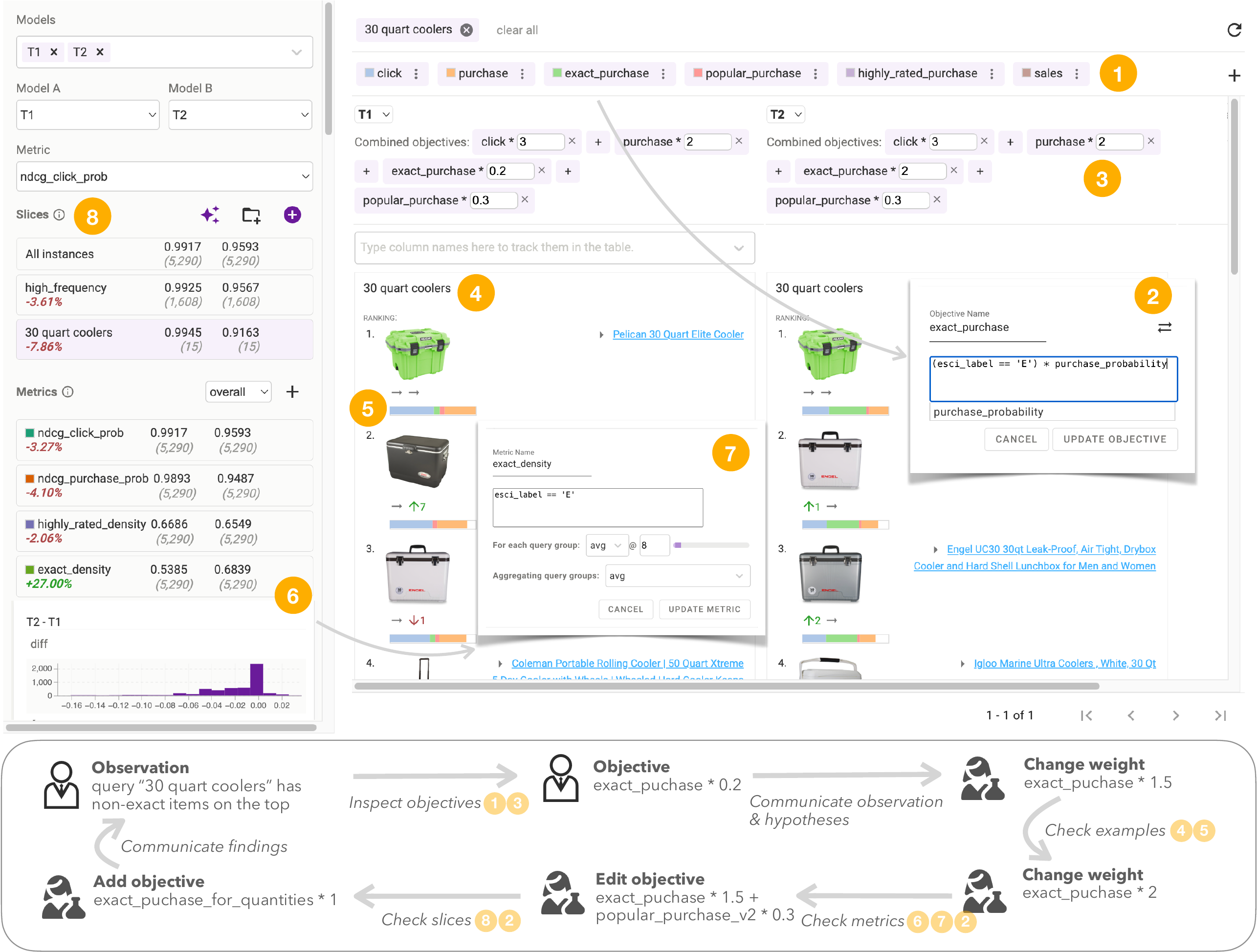}
    \caption{\toolname{}'s interface and example usage. 
    \toolname{} surfaces objectives as the first class citizen in \circleone{} objective overview bar, and allows users to interactively \circletwo{} inspect, edit, or create objectives.
    Users can \circlethree{} specify how multiple objectives are combined and incorporated into a model, and observe the impact in real-time.
    Users can look at \circlefour{} side-by-side comparison for example-level information, and \circlefive{} tie rankings back to objectives for explanations when needed.
    Users can also look at \circlesix{} metrics and \circleeight{} slices for aggregated information, with the ability to interactively define \circleseven{} new metrics and new slices, and \circlesix{} inspect slices with larger metric differences.
    Below the interface, we demonstrate how in our running examples, stakeholders can use \toolname{} to translate observations to actionable feedback, explore different designs, and gather evaluation information.
    }
    \label{fig:overview}
\end{figure*}

\section{\toolname}
\label{sec:definition}

We proposed \toolname{}, a conceptual framework for multi-objective ranker design and evaluation,
surfacing \textit{objective} as the core concept (Figure~\ref{fig:concept}).

\paragraph{What are objectives?}
Objectives can be formally defined as \textit{``functions we want to minimize or maximize''} for machine learning~\cite{Goodfellow-et-al-2016}.
For example, an objective for user click can be defined as a cross-entropy loss function:

\[
\mathcal{L}_{\text{click}} = - \frac{1}{N} \sum_{i=1}^{N} \left[ r_{ui} \log(\hat{r}_{ui}) + (1 - r_{ui}) \log(1 - \hat{r}_{ui}) \right]
\]

where \( r_{ui} \) is whether user \( u \) clicked item \( i \), \( \hat{r}_{ui} \) is the softmax predicted click \textit{probability} for the same user and item, and \( N \) is the total number of user-item pairs in the dataset.
Model training is expected to optimize for this objective, i.e., minimizing the gaps between predictions and ground truth.
In reality, there are usually multiple objectives (e.g., click, purchase, relevance, ratings) that can be defined and optimized for, which requires practitioners to actively explore different designs and trade-offs. 

For evaluation, practitioners need to design appropriate metrics for each objective. 
This is usually a one-to-many mapping, with different metrics capturing different notions of the same objective: 
For example, MAP~\citep{zhu2004recall} and NDCG~\citep{jarvelin2002cumulated} can both be used for the user click objective here to measure the ``goodness'' of a produced ranking in terms of clicks, with NDCG capturing additional position information.

Objectives can also be interpreted colloquially and connected to stakeholder feedback and concrete examples:
For example, the click objective captures user click information and can be held accountable when stakeholders identify examples where clickbait issues~\cite{wang2021clicks} promote attractive-looking yet low-quality items.
Effectively, objectives can serve as a boundary object~\cite{boundary-object} between stakeholders.

\paragraph{Objective-centered ranker design}
We argue that multi-objective ranker design should be objective-centered,
because objectives can serve as a boundary object for communication, 
help design space navigation (cf. Section~\ref{sec:design-space-nav}), and help information foraging for evaluation (cf. Section~\ref{sec:information-foraging}).

We implemented \toolname{} as an interactive system (Figure~\ref{fig:overview}) based on Zeno~\cite{cabrera2023zeno}, a framework that supports interactive behavioral evaluation.
\new{The system was implemented and iterated over three months, with weekly feedback from internal stakeholders.}
To walk through how \toolname{} implements the conceptual framework, we use the following running example:

\aptLtoX[graphic=no,type=html]{\begin{framed}
{Cynthia is an ML engineer responsible for an E-commerce ranking model, which takes in a query and a list of items and returns a ranked list. 
Currently, the model is optimized for four objectives:
\objective{click} for estimated click probability,
\objective{purchase} for estimated purchase probability,
\objective{exact\_purchase} for textual relevance (i.e., whether an item is an exact match) weighted by purchase,
and \objective{popular\_purchase} for popularity (i.e., units sold) weighted by purchase.
Cynthia is working with a product manager, Eric, to decide whether to incorporate his feedback into the next model iteration.}
\end{framed}}{\quo{Cynthia is an ML engineer responsible for an E-commerce ranking model, which takes in a query and a list of items and returns a ranked list. 
Currently, the model is optimized for four objectives:
\objective{click} for estimated click probability,
\objective{purchase} for estimated purchase probability,
\objective{exact\_purchase} for textual relevance (i.e., whether an item is an exact match) weighted by purchase,
and \objective{popular\_purchase} for popularity (i.e., units sold) weighted by purchase.
Cynthia is working with a product manager, Eric, to decide whether to incorporate his feedback into the next model iteration.}}

\subsection{Objective as Boundary Object}

First, \toolname{} explicitly surfaces objectives in the \circleone{} \textbf{objective overview bar} (\textbf{G1}) so that stakeholders can have a shared understanding of what objectives are currently used.

\aptLtoX[graphic=no,type=html]{\begin{framed}
Eric observes that in a query ``30 quart coolers'', the current model promotes too many items that are not exact matches in terms of textual relevance (i.e., different sizes), even though users are actively seeking items with a specific size. 
Eric notices that the model is currently optimized for four objectives: \objective{click}, \objective{purchase}, \objective{exact\_purchase}, and \objective{popular\_purchase}.
\end{framed}}{\quo{Eric observes that in a query ``30 quart coolers'', the current model promotes too many items that are not exact matches in terms of textual relevance (i.e., different sizes), even though users are actively seeking items with a specific size. 
Eric notices that the model is currently optimized for four objectives: \objective{click}, \objective{purchase}, \objective{exact\_purchase}, and \objective{popular\_purchase}.}}

Furthermore, \toolname{} supports users to inspect the exact definition of each objective. 
This helps stakeholders locate objectives relevant to their observations.

\aptLtoX[graphic=no,type=html]{\begin{framed}Eric finds the objective most relevant to his observations is \objective{exact\_purchase}, defined as {\texttt{esci\_label} $==$ \texttt{`E'}}, where \texttt{esci\_label} is a feature annotating whether an item is an exact match, substitute, complement, or irrelevant to the current query.\end{framed}}{\quo{Eric finds the objective most relevant to his observations is \objective{exact\_purchase}, defined as {$\texttt{esci\_label} == \texttt{`E'}$}, where \texttt{esci\_label} is a feature annotating whether an item is an exact match, substitute, complement, or irrelevant to the current query.}}

\toolname{} is also explicit about how objectives are incorporated into the model with \circlethree{} \textbf{model definitions}  (\textbf{G1}), 
helping stakeholders provide constructive feedback grounded in existing model designs.

\aptLtoX[graphic=no,type=html]{\begin{framed}Eric finds the current model is trained using a linear combination of objectives:
\text{\objective{click}} * 3 + \text{\objective{purchase}} * 2 + \text{\objective{exact\_purchase}} *0.2 + \text{\objective{popular\_purchase}} *0.3. Eric notices that \objective{exact\_purchase} has a weight of 0.2 -- he suspects that the current weight is too small to promote all items that are exact matches.
Eric communicates this, along with his concrete observations, to Cynthia.\end{framed}}{\quo{Eric finds the current model is trained using a linear combination of objectives:
\text{\objective{click}} * 3 + \text{\objective{purchase}} * 2 + \text{\objective{exact\_purchase}} *0.2 + \text{\objective{popular\_purchase}} *0.3. Eric notices that \objective{exact\_purchase} has a weight of 0.2 -- he suspects that the current weight is too small to promote all items that are exact matches.
Eric communicates this, along with his concrete observations, to Cynthia.}}

\looseness=-1
Our example here shows one way to incorporate multiple objectives through pre-training linear aggregation~\cite{dong2010towards}.
There are many other ways objectives can be incorporated into the model, from post-training aggregation~\cite{zhao2019recommending} to heuristic reranking over existing models~\cite{chen2023controllable} (cf. Section~\ref{sec:moml} for a more detailed discussion), 
but the core design of \toolname{} stays the same regardless of the training methods.

\subsection{Objectives to Support Design Space Navigation}
\label{sec:design-space-nav}

\toolname{} not only helps stakeholders provide more constructive feedback on ranker design but also makes the design process itself much easier by helping them navigate the design space.
This is supported both by \circleone{} \textbf{objective overview bar}, which provides the design ingredients, and by \circlethree{} \textbf{model definitions}, 
which shows how multiple objectives are combined and further allows users to update the aggregation methods with both smaller changes (weight-tuning) and bigger changes (adding or removing objectives).

\aptLtoX[graphic=no,type=html]{\begin{framed}Receiving Eric's feedback, Cynthia decides to update the weight of \objective{exact\_purchase} to 1.5 to see if this could help and what the trade-off is. 
She also tests out a few different designs, including trying different weights of \objective{exact\_purchase} and changing thresholds for \objective{popular\_purchase}.\end{framed}}{\quo{Receiving Eric's feedback, Cynthia decides to update the weight of \objective{exact\_purchase} to 1.5 to see if this could help and what the trade-off is. 
She also tests out a few different designs, including trying different weights of \objective{exact\_purchase} and changing thresholds for \objective{popular\_purchase}.
}}

\toolname{} encourages users to think about design space exploration in terms of objectives, 
and helps them track what they have explored and brainstorm what to explore next.
Furthermore, \toolname{} supports \circletwo{} \textbf{interactive objective edits and definitions},
such that users can easily manipulate the objective space (\textbf{G3}).
This is particularly helpful when they want to experiment with new objectives.

\aptLtoX[graphic=no,type=html]{\begin{framed}Motivated by the evaluation results, Cynthia decides relevance is particularly important for queries with queries with quantity information, where users likely target exact items only.
She defines and adds an additional objective \objective{exact\_purchase\_for\_quantities} that selectively applies \objective{exact\_purchase} to a data slice.\end{framed}}{\quo{Motivated by the evaluation results, Cynthia decides relevance is particularly important for queries with queries with quantity information, where users likely target exact items only.
She defines and adds an additional objective \objective{exact\_purchase\_for\_quantities} that selectively applies \objective{exact\_purchase} to a data slice.}}

\subsection{Objective to Support Information Foraging in Evaluation}
\label{sec:information-foraging}
Each model design users explore needs to be evaluated.
\toolname{} highlights a comprehensive set of evaluation information (\textbf{G2}) and supports effective information foraging~\cite{information-foraging} with objectives (\textbf{G1}).

Most noticeably, \toolname{} helps users forage useful information at example-level with a \circlefour{} \textbf{side-by-side comparison} view of examples,
where users can visually inspect concrete ranking results with rank differences highlighted.
Users can also find additional information by tracking additional (objective-related) columns, or expanding a specific item to inspect details (\textbf{G2}).

\aptLtoX[graphic=no,type=html]{\begin{framed}After updating the objectives, Cynthia checks the examples associated with ``30 quart coolers.'' 
She finds that this change successfully demotes two non-exact items on the top.\end{framed}}{\quo{
After updating the objectives, Cynthia checks the examples associated with ``30 quart coolers.'' 
She finds that this change successfully demotes two non-exact items on the top.
}}

\toolname{} explicitly tie example rankings back to objectives, with an 
\circlefive{} \textbf{objective attribution chart} under each item  (\textbf{G1}).
The attribution chart visualizes how much each objective contributes to a specific item's ranking and can be computed using model explanability techniques~\cite[e.g.,][]{ribeiro2016should, baehrens2010explain}.
Together with objective-related features, objective attribution charts help users identify information that is useful for model iteration.
In other words, \toolname{} helps enhance ``scent''~\cite{information-foraging} of relevant information for more efficient evaluation.

\aptLtoX[graphic=no,type=html]{\begin{framed}Cynthia looks at the example and finds the two non-exact items are demoted specifically because the objective of \objective{exact\_purchase} (green in the charts) takes a much bigger portion in the combined objective.
She also checks the promoted items, noticing they have medium-level review counts.\end{framed}}{\quo{Cynthia looks at the example and finds the two non-exact items are demoted specifically because the objective of \objective{exact\_purchase} (green in the charts) takes a much bigger portion in the combined objective.
She also checks the promoted items, noticing they have medium-level review counts.
}}

Beyond examples, \toolname{} also features a \circlesix{} \textbf{metric panel}, which provides an overview over all metrics under tracking. 
The metrics can usually be derived from objective, though there is not necessarily a one-to-one mapping.
In \toolname{}, users can easily glance over the metric changes at the dataset, slice, or example level, as well as zoom into specific slices with larger metric differences (\textbf{G2}).

\aptLtoX[graphic=no,type=html]{\begin{framed}In the current iteration, there are four metrics under tracking: \metric{ndcg\_click\_prob} and \metric{ndcg\_purchase\_prob} measure how well highly clicked (purchased) items are placed on the top, with a position decay~\cite{jarvelin2002cumulated}, 
while  \metric{exact\_density} and \metric{highly\_rated\_density} measure how many exact match or highly rated items are present in top-8.

\noindent Cynthia notices that, even though boosting \objective{exact\_purchase} indeed promotes textually relevant items to the top (hence higher \metric{exact\_density}), other key metrics like \metric{ndcg\_purchase\_prob} and \metric{highly\_rated\_density} can drop a lot over the entire dataset.
Cynthia tests out a few different designs to find one that balances \metric{ndcg\_purchase\_prob} and \metric{exact\_density} relatively well.
\end{framed}}{\quo{In the current iteration, there are four metrics under tracking: \metric{ndcg\_click\_prob} and \metric{ndcg\_purchase\_prob} measure how well highly clicked (purchased) items are placed on the top, with a position decay~\cite{jarvelin2002cumulated}, 
while  \metric{exact\_density} and \metric{highly\_rated\_density} measure how many exact match or highly rated items are present in top-8.

\noindent Cynthia notices that, even though boosting \objective{exact\_purchase} indeed promotes textually relevant items to the top (hence higher \metric{exact\_density}), other key metrics like \metric{ndcg\_purchase\_prob} and \metric{highly\_rated\_density} can drop a lot over the entire dataset.
Cynthia tests out a few different designs to find one that balances \metric{ndcg\_purchase\_prob} and \metric{exact\_density} relatively well.
}}

More importantly, users can customize any metrics they find useful with \circleseven{} \textbf{metric definitions}, and interactively track them.
If users have specific hypotheses, they can even define data slices and track their changes with \circleeight{} \textbf{interactive slicing} (\textbf{G3}) .

\aptLtoX[graphic=no,type=html]{\begin{framed}
Cynthia further investigates into data slices where the metric \metric{ndcg\_purchase\_prob} drops most.
She notices that for some exploratory user queries that are broad, ambiguous, or even misleading,  
there is a strong conflict between textual relevance and user purchases,
In these cases, users tend to buy a lot of supplementary items,
but the new objectives can downrank these items a lot despite frequent user purchases.

\noindent Motivated by this observation, Cynthia defines a slice on queries with quantities, where users likely target only for exact items -- she finds the model performs well on this slice with new objectives.
She further iterates the model design and tests out different designs until she finds a few satisfactory.
Cynthia decides to move on with the identified designs, communicates her exploration and analysis back to Eric, and starts a few model training sessions.
\end{framed}}{\quo{Cynthia further investigates into data slices where the metric \metric{ndcg\_purchase\_prob} drops most.
She notices that for some exploratory user queries that are broad, ambiguous, or even misleading,  
there is a strong conflict between textual relevance and user purchases,
In these cases, users tend to buy a lot of supplementary items,
but the new objectives can downrank these items a lot despite frequent user purchases.

\noindent Motivated by this observation, Cynthia defines a slice on queries with quantities, where users likely target only for exact items -- she finds the model performs well on this slice with new objectives.
She further iterates the model design and tests out different designs until she finds a few satisfactory.
Cynthia decides to move on with the identified designs, communicates her exploration and analysis back to Eric, and starts a few model training sessions.
}}

\section{Evaluation}

To evaluate \toolname{}, we want to understand how it supports ranker design.
As discussed before, ranker design is a wicked problem that requires thinking about and trading off multiple objectives -- there is no single success criteria or quality measure,
hence, assessing the speed of task completion would be an inadequate measure because it is easy to create a poorly thought out solution quickly with or without tool support.
Instead, we want to evaluate how much users explore and evaluate trade-offs, 
as we consider the depth of engagement as a proxy for their efforts put into creating a well-thought-out and balanced solution. 
While we do not have any ground truth to evaluate the quality of a solution (which is hardly ever possible for a wicked problem), 
we can measure how deeply users engage with reasoning about the problem in a given time.
We expected \toolname{} can support users to explore design space more broadly, conduct more comprehensive evaluations for decision-making, and derive better justifications considering trade-offs.

More specifically, we conducted a user study to evaluate:
\begin{itemize}
    \item \textbf{RQ1 (more efficient navigation)}: To what degree does \toolname{} help users explore the design space more easily and efficiently? 
    \item \textbf{RQ2 (more informed decision-making)}: How well does \toolname{} help users make more informed decision? 
    \item \textbf{RQ3 (more trade-off thinking)}: How well does \toolname{} encourage users to think about and communicate trade-offs? 
\end{itemize}

\subsection{Study Design}

\subsubsection{Experimental conditions}
We design our user study as a within-subject controlled experiment,
where participants complete two tasks in two conditions: \textit{treatment} and \textit{control}.
In the treatment condition, participants use \toolname{}, 
while in the control condition, participants use Jupyter notebooks, as commonly used in their existing workflow.
To make it a fair comparison, we also provide additional utility functions for designing objectives and computing metrics in the control condition, such that the control group is better supported than an average practitioner doing this task.

\subsubsection{Procedure}
We conducted the study one-on-one with all participants. Each session lasted for 90 minutes and had a structure as follows:
The participants first filled out a pre-study survey for demographics and expertise information.
Next, the participants went through an interactive tutorial in Jupyter Notebook, where they were introduced to (1) the dataset used in the study, (2) provided notebook utilities for the control condition, and (3) key functionalities of \toolname{}.
The tutorial serves to equip participants with background information for the study tasks.
After the tutorial, the participants were asked to try a demo task with \toolname{}, to make sure the participant understood the task and how to use \toolname{}.
The introductory part took up to 30 minutes.

Next, participants worked on the two tasks for 25 minutes each, one in the treatment condition and one in the control condition. 
The participants were asked to work on the tasks think-aloud~\cite{holtzblatt1997contextual}, such that we could better understand their thought processes and decision points.
To mitigate learning effects, we use a Latin square design~\cite{box_2009} with four groups, counterbalancing (1) which condition a participant encounters first, and (2) which task a participant works on first. 
In the end, participants filled out a post-study survey for their feedback (details in Appendix~\ref{sec:survey}).

\subsubsection{Tasks}
We designed the tasks in the following structure:
The participant was first shown feedback on specific model outputs from other (hypothetical) stakeholders. 
They then had 20 minutes to explore, evaluate, and analyze different model designs think-aloud to understand (1) whether and how the feedback can be incorporated and (2) what are the potential trade-offs.
Finally, they had 5 minutes to draft a response to the stakeholder feedback based on their exploration and analysis.

For our user study, we designed two task scenarios with similar difficulty, using the public ESCI dataset~\cite{reddy2022shopping}.\footnote{We annotated the dataset with additional synthetic features, ending up with having \textit{text relevance, click-through probability, purchase probability, review ratings, review counts, units sold} as objective-relevant columns.}

\begin{enumerate}
    \item A stakeholder found the query ``30 quart coolers'' has lots of products with different sizes (e.g., 54 quarts) on the top. These products are not exact matches and might be irrelevant to customers looking for coolers with a specific size.
    \item A stakeholder found the query ``uconn hoodie'' has lots of products with bad ratings on the top. These products are poorly sold and rated, and can negatively impact customers' shopping experience.
\end{enumerate}

\subsubsection{Measurements and analysis}

\paragraph{User activity}
We characterize user activities into two categories: design and evaluation. 
For design, we distinguish between small-step exploration (weight-tuning) and big-step exploration (others), to understand how \toolname{} impacts users' design exploration in more nuances.
For evaluation, we further break it down into example-based and metric-based evaluations, and distinguish between standard evaluations (dataset-level metrics, provided anecdotes) and additional evaluations (others), to understand how \toolname{} impact users' information-seeking behaviors.
We re-construct participants' activity sequences from the telemetry data (for treatment) and execution history (for control) collected during their interactions, 
and one author went through all screen recordings to validate the activity sequences. 
The final user activity sequences produced are visualized in Figure~\ref{fig:activity}.

For {RQ1} (more efficient navigation), we measure how much design space users explore, with \textbf{distinct trade-offs (M1)} users explore, which we define as the number of different objectives users design and evaluate.
We also measure how many design dimensions users explore, with \textbf{distinct big-step trade-offs (M2)}, where users change objective constituents.
To understand the complexity of objectives users explore, we additionally measure {degree of feature interaction} for each new or edited objective.

For {RQ2} (more informed decision-making), we measure how comprehensive evaluations users conduct for each trade-off, with \textbf{distinct evaluations per trade-off (M3)}. More comprehensive evaluations imply more informed decision-making.
We consider each evaluation result that gives new information as distinct -- 
for example, if users look at the same example multiple times for one trade-off, we would consider them as one distinct evaluation.

We further measure the comprehensiveness of users' \textit{overall} evaluation, with \textbf{distinct additional evaluations (M4)}, to understand how much users go beyond standard setups of dataset metrics and provided anecdotes.
We also measure how balanced users' evaluations are, with \textbf{metric-example balance (M5)}, to understand how much users rely on one-sided information.
We define this metric as KL-divergence from the uniform distribution:

\[
   D_{\text{KL}}(Q \parallel P) = Q(e) \log \left(\frac{Q(e)}{P(e)}\right) + Q(m) \log \left(\frac{Q(m)}{P(m)}\right)
\]

where $P(e)=P(m)=\frac{1}{2}$, and $Q(e)$ ($Q(m)$) measures the proportion of example-based (metric-based) evaluations.
The smaller the metric, the more balanced users' evaluations.

\begin{figure}[t]
    \centering
    \includegraphics[width=0.85\linewidth]{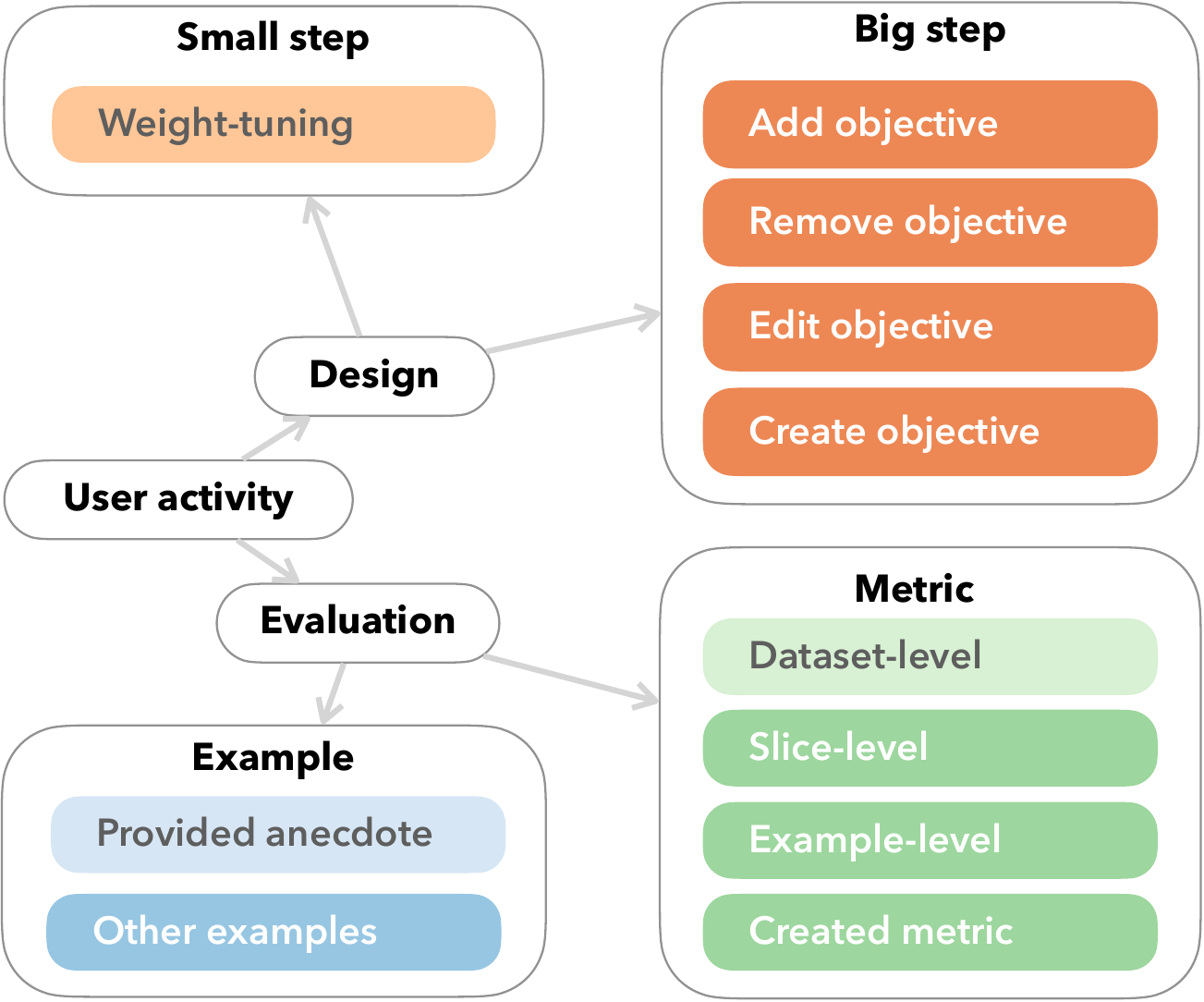}
    \caption{Taxonomy of user activities: We characterize user activities into two categories: design and evaluation. 
    For design, we distinguish between small-step exploration (weight-tuning) and big-step exploration (others), to understand how \toolname{} impacts users' design exploration in more nuances.
    For evaluation, we further break it down into example-based and metric-based evaluations, and distinguish between standard evaluations (dataset-level metrics, provided anecdotes) and additional evaluations (others), to understand how \toolname{} impact users' information-seeking behaviors.
    }
    \label{fig:activity-breakdown}
\end{figure}

We summarize all our measurements collected from user activity in Table~\ref{tab:metrics}.
For all these measurements, we also analyzed with a repeated measures ANOVA analysis, 
testing how much the condition (whether participants use \toolname{}) impacts the measurements, 
while considering the potential impact from other independent variables: 
In our analysis, we test to what degree our tool, the task, the order (tool first or tool last), and the participant's past experiences explain variance in the outcome variables (cf. Table~\ref{tab:anova}).

\begin{table*}[t]
\centering

\small
\begin{tabular}{>{\raggedleft\arraybackslash}p{2.5cm}|p{4.5cm} p{3cm} p{3.5cm} }
\toprule
    \textbf{Metric} & \textbf{Definition} & \textbf{Hypothesis}  & \textbf{Result} \\
        \midrule
        \textbf{distinct trade-offs (M1)}      &  The number of different objectives users design and evaluate & Increased trade-off exploration with \toolname{} & \vtop{\vskip-8pt\hbox{\includegraphics[width=\linewidth]{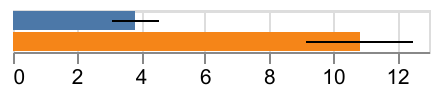}}}       \\ 
        \textbf{distinct big-step trade-offs (M2)}       & The number of different objectives users design and evaluate, where objectives are added, removed, or edited & Increased big-step trade-off exploration with \toolname{}    & \vtop{\vskip-8pt\hbox{\includegraphics[width=\linewidth]{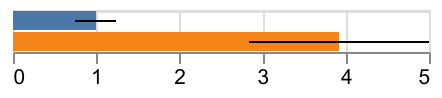}}}     \\ 
         \\
        \textbf{distinct evaluations per trade-offs (M3)}                 & The number of different evaluations users conduct for each trade-off & Increased evaluation per exploration with \toolname{}     & \vtop{\vskip-8pt\hbox{\includegraphics[width=\linewidth]{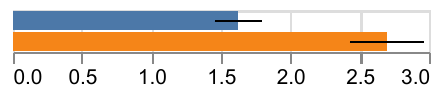}}}        \\ 
        \textbf{distinct additional evaluations (M4)}       & The number of different additional metric-based or example-based evaluations users conduct & Increased additional evaluation with \toolname{}    & \vtop{\vskip-8pt\hbox{\includegraphics[width=\linewidth]{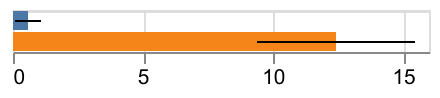}}}        \\ 
         \\
        \textbf{metric-example balance (M5)}       &  KL-divergence from the uniform distribution between metric-based and example-based evaluations  & More balanced evaluation (\textbf{M5} $\downarrow$) with \toolname{}  & \vtop{\vskip-8pt\hbox{\includegraphics[width=\linewidth]{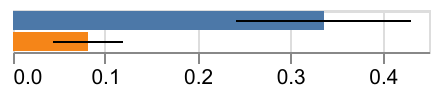}}}        \\ 
        \bottomrule
       \multicolumn{3}{l}{} & \multicolumn{1}{r}{\aptLtoX[graphic=no,type=html]{\textcolor{orange}{$\blacksquare$}}{\treatment{}} treatment \quad \aptLtoX[graphic=no,type=html]{\textcolor{blue}{$\blacksquare$}}{\control{}} control }
\end{tabular}
\caption{User study metrics and results for RQ1 and RQ2. With \toolname{}, participants explored more {distinct} trade-offs (\textbf{M1}) in {bigger} steps (\textbf{M2}). They also conducted more {distinct} evaluation (\textbf{M3}) beyond standard setups (\textbf{M4}) in a more {balanced} way (\textbf{M5}).
}
\label{tab:metrics}
\end{table*}

\paragraph{User responses}
For {RQ3} (more trade-off thinking), we annotated users' responses to stakeholders on whether they directly mention trade-offs.
We classified any responses mentioning tensions between different objectives (metrics) as directly mentions (e.g., \textit{``achieve better [metricA] without compromising [metricB]''}).
We measure the proportion of responses mentioning trade-offs.
In addition, we went through user study transcripts, and identified and counted all verbal mentions of trade-offs.

For our annotations, we have two authors independently annotate a same subset of responses, achieving substantial agreement (0.67 kappa score). One author proceeded with the rest of annotations.

\paragraph{Survey}
For all our findings, we also triangulate the findings using users' survey responses,
where they rate different aspects of each study condition (trade-off understanding, usability, usefulness, etc.) and the importance of different \toolname{} features (e.g., side-by-side visualization, metric tracking).
We also quote their written feedback for the relevant findings.

\subsubsection{Participants}
We recruited 12 participants from a large tech company. 
All participants have prior ML experience and currently work on product ranking, with 83\% of them ``extremely familiar'' or ``very familiar'' with Jupyter notebooks in self-report.
\new{Their daily work areas span across feature extraction, model training, anomaly detection, offline model evaluation, online A/B testing, etc.}
As is the standard practice, we pilot-tested the evaluation with 4 participants from the same company, which are not included in the final results.

\begin{table*}[t]
\centering

\small
\begin{tabular}{r|p{1.75cm} p{2.25cm} p{2.75cm} p{2.55cm} p{2.15cm}}
\toprule
    & \textbf{distinct trade-offs (M1)} & \textbf{distinct big-step trade-offs (M2)} & \textbf{distinct evaluations per trade-offs (M3)} & \textbf{distinct additional evaluations (M4)} & \textbf{metric-example balance (M5)} \\
        \midrule
        \textbf{Interv.: Used \toolname{}?}      & 23.81***  & 8.18*   & 11.03**     & 12.16**   & 6.50*      \\ 
        \textbf{Task number}                    & 7.14*     & 6.42*  & 2.71     & 0.12      & 3.14     \\ 
        \textbf{Tool Order}                 & 0.34      & 0.54      & 0.43      & 0.01      & 0.02      \\ 
        \textbf{Notebook experience}       & 4.97*     & 0.42      & 0.19      & 0.04      & 0.79      \\ 
        \bottomrule
      \multicolumn{3}{l}{} & \multicolumn{3}{r}{\small{$^{***}p<0.001,\quad ^{**}p<0.01,\quad ^{*}p<0.05$}}
\end{tabular}
\caption{User study ANOVA results: We report the F-value and p-value, which quantify the extent to which each variable accounts for the observed variances. Our analysis reveals that the use of \toolname{} significantly explains the differences for all measurements with the biggest impact, while all other variables can not significantly explain the observed variances, except for Task number for \textbf{M1} and \textbf{M2}, and notebook experience for \textbf{M1}, with smaller F-values.
}
\label{tab:anova}
\end{table*}

\subsubsection{Limitations}
Our user study is designed as a controlled experiment. 
Controlled experiments give us the power to ensure high confidence in the reliability of the findings in the given context with statistical techniques,
but the results might not generalize easily to other tasks, settings, or ML practitioners beyond our participant population.
This is a common trade-off when designing evaluations~\cite{siegmund2015views} -- readers should be careful when generalizing findings beyond our study setting.
\new{In addition, the controlled experiments have a short study duration where \toolname{} is introduced to participants for the first time, leading to potential novelty effects~\cite{koch2018novelty}.}

Our analysis uses a series of metrics as proxies to measure how users explore design space, conduct evaluation, and think about trade-offs.
We do not have a single metric to measure the ``goodness'' of the derived solutions, as this would require a fixed view of how to prioritize different objectives.
Our analysis also relies on human annotations for some metrics, which can be inherently subjective and unreliable -- we mitigated the problem with multiple raters to establish annotation reliability.

\aptLtoX[graphic=no,type=html]{\begin{figure*}
    \centering
    \includegraphics[width=\linewidth]{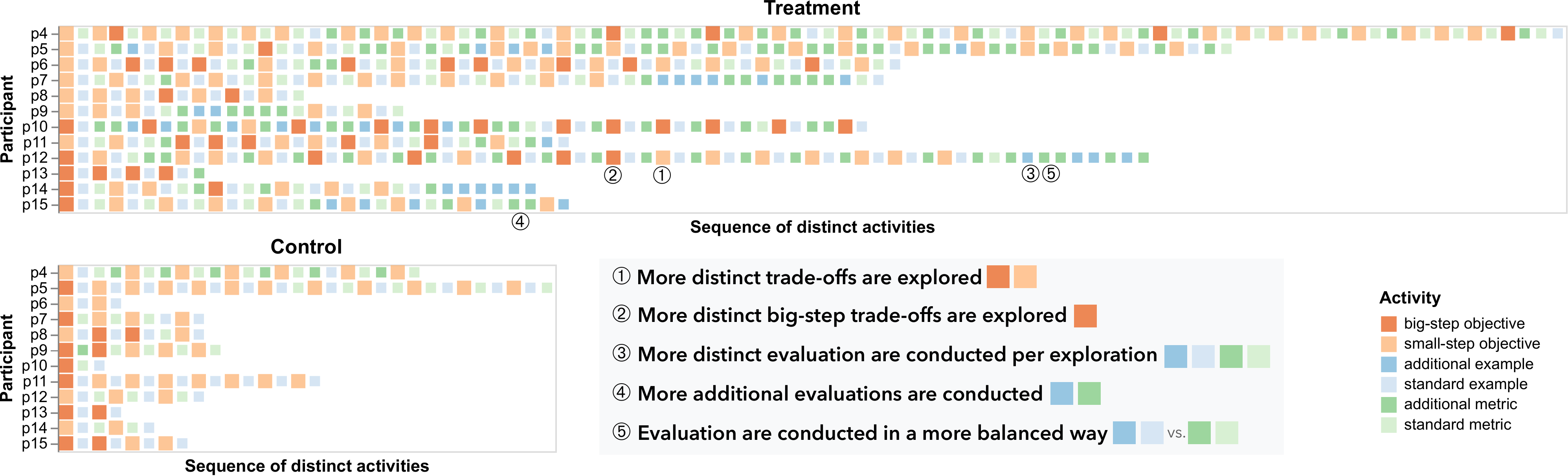}
    \caption{Participant's sequence of distinct activities.
    With \toolname{}, participants explored more \textit{distinct} trade-offs {\textcolor{deeporange}{$\blacksquare$}} {\textcolor{lightorange}{$\blacksquare$}}, in \textit{bigger} steps {\textcolor{deeporange}{$\blacksquare$}}, and conducted more \textit{distinct} evaluation {\textcolor{deepblue}{$\blacksquare$}} {\textcolor{lightblue}{$\blacksquare$}} {\textcolor{lightblue}{$\blacksquare$}} {\textcolor{deepgreen}{$\blacksquare$}} {\textcolor{lightgreen}{$\blacksquare$}} beyond standard setups ({\textcolor{deepblue}{$\blacksquare$}} {\textcolor{deepgreen}{$\blacksquare$}} vs. {\textcolor{lightblue}{$\blacksquare$}} {\textcolor{lightgreen}{$\blacksquare$}}) in a more \textit{balanced} way ({\textcolor{deepblue}{$\blacksquare$}} {\textcolor{lightblue}{$\blacksquare$}} vs. {\textcolor{deepgreen}{$\blacksquare$}} {\textcolor{lightgreen}{$\blacksquare$}}). 
    Overall participants also explored big-step changes throughout the session with \toolname{} (vs. mostly only did big-step changes in the beginning followed by small weight-tuning when using notebooks).
}
    \label{fig:activity}
\end{figure*}}{\begin{figure*}[t]
    \centering
    \includegraphics[width=\linewidth]{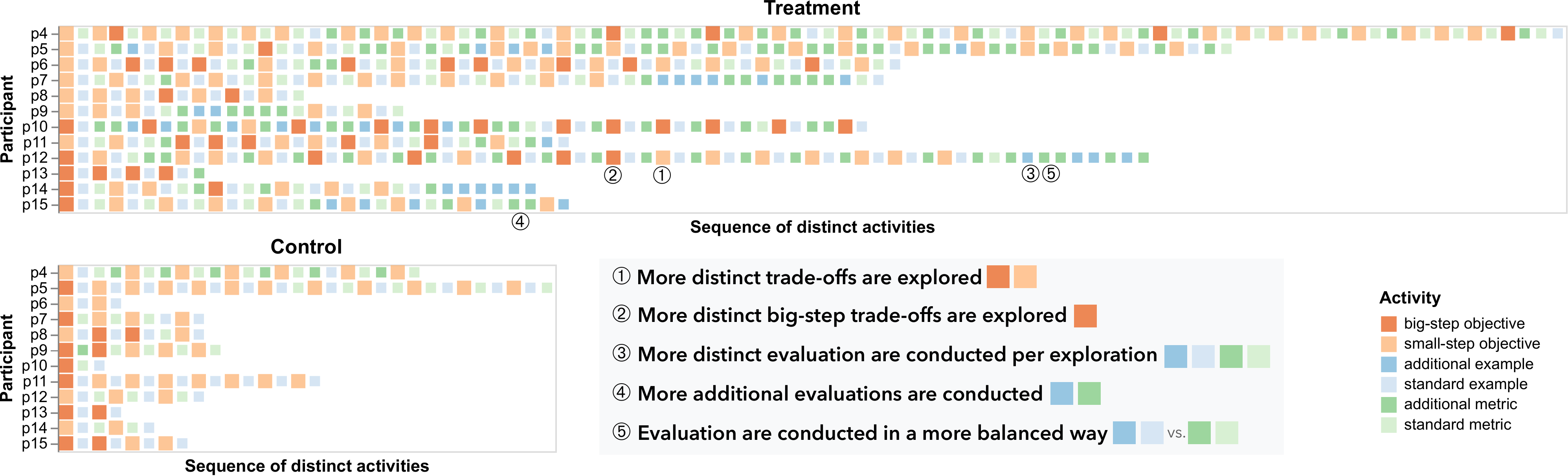}
    \caption{Participant's sequence of distinct activities.
    With \toolname{}, participants explored more \textit{distinct} trade-offs \bigstep\smallstep, in \textit{bigger} steps \bigstep, and conducted more \textit{distinct} evaluation \addexm\stdexm\addmetric\stdmetric{} beyond standard setups (\addexm\addmetric vs. \stdexm\stdmetric) in a more \textit{balanced} way (\addexm\stdexm vs. \addmetric\stdmetric). 
    Overall participants also explored big-step changes throughout the session with \toolname{} (vs. mostly only did big-step changes in the beginning followed by small weight-tuning when using notebooks).
}
    \label{fig:activity}
\end{figure*}}

\subsection{Finding: \toolname{} Helps Users Explore Design Space More Efficiently (RQ1)}

\subsubsection{Users explored 183\% more distinct trade-offs with \toolname{}}
We found that on average, users can explore much more \textit{distinct} trade-offs (\textbf{M1}) with \toolname{} (10.8 vs. 3.8 in control, Table~\ref{tab:metrics}).
This \nobreak demonstrates that, given the same amount of time, users can explore the design space more efficiently,
correlating with user perception (Table~\ref{tab:ratings}) that \toolname{} is easier to use (83\% vs. 41\% for notebooks), and helps them accomplish the tasks more easily (91\% vs. 33\% for notebooks).

A closer inspection of screen recordings reveals that, in the control condition, users spend much more time foraging example-level information, while in the treatment condition, the side-by-side visualization makes it much faster to gather the same information.
This echoes their own perception that side-by-side comparison is the most important feature in \toolname{}: 100\% participants rate it as ``extremely important'' or ``very important'' (Figure~\ref{fig:importance}),
as side-by-side comparison \participantQuote{provides a more visual way to experiment with tradeoffs}{4}.

\subsubsection{Users explored 292\% more trade-offs in larger steps with \toolname{}}
Further breaking down the trade-offs users explored, we found that users not only explored more distinct trade-offs but also explored the design space in larger steps:
Indeed, they explored 291.7\% more \textit{big-step} changes (\textbf{M2}, 3.9 vs. 1.0 in control,  Table~\ref{tab:metrics}) and created or edited 142.9\% more objectives (1.4 vs. 0.6 in control).
This is well illustrated by the activities of P11 (shown in Figure~\ref{fig:activity}): 
In the control condition,  P11 only defined and added a new objective once, with the remaining time exclusively focusing on weight-tuning,
while in the treatment condition, P11 defined, added and edited objectives throughout the process.

Comparing the objectives users defined, \toolname{} also enabled users to explore more complex objectives that they would not have considered before:
Treatment group users are observed to explicitly explore feature interactions (e.g., \texttt{(esci\_label == `E') $\cdot$ purchase\_probability $\cdot$ (review\_rating > 4)}, P10), 
while control group users exclusively create simple objectives (e.g., \texttt{esci\_label == `E'}).
Overall, treatment groups defined or edited 19 objectives with 1.6 interactions on average, 
while control groups only defined or edited 7 objectives with 1.1 interactions on average.

Combining the results, we found that \toolname{} helps users explore design space more efficiently, and in the way that they also explore bigger changes and more complex interactions.
That is, given the same amount of time, users are able to test out more \textit{divergent} design ideas with \toolname{}.

\subsection{Finding: \toolname{} Leads to More Informed Decision Making (RQ2)}
\label{sec:rq2}

Beyond more efficient exploration of the design space, we also observed that users on average conducted 65.7\% more distinct evaluations per exploration with \toolname{} (\textbf{M3}, Table~\ref{tab:metrics}).
Participants felt that \toolname{} helps them  \participantQuote{easy to process (information) and estimate the effects of changes quickly}{15} and mentally \participantQuote{makes it easier to navigate through tradeoffs}{12}.

Breaking down the evaluations, we found that with \toolname{}, users are especially encouraged to explore additional evaluations (\textbf{M4}, 12.4 vs. 0.6,  Table~\ref{tab:metrics}), while they almost exclusively focused on standard setups (except for P4) in the control conditions.
For example, P12 only checked evaluation results on overall metrics and the provided task example when using notebooks, but conducted a much more extensive evaluation including additional slice-level metrics and additional examples with \toolname{} (Figure~\ref{fig:activity}).
This shows that \toolname{} not only encourages more frequent evaluation, but also encourages evaluation beyond overall aggregated metrics and fosters generalizability thinking (additional examples).

Meanwhile, we also observed that users are conducting evaluations in a more balanced (\textbf{M5}, 75.8\% closer to uniform,  Table~\ref{tab:metrics}) way:
In the control conditions, some participants almost exclusively focus on one kind of evaluation -- metric-based (e.g., P9) or example-based (e.g., P15).
In contrast,  with \toolname{}, participants are more likely to think about both metrics and examples at the same time, because they can easily check \participantQuote{if the solution works by looking at the (3) metrics block and (4) side by side block, both updated automatically}{4}, with \participantQuote{less mental load for human}{14}.

Additionally, we observed that users are also more likely (+200\%) to define new metrics with \toolname{}.
For example, P4 added new metrics to capture different definitions of \objective{popularity} with different thresholds for \texttt{review\_rating} to decide what counts as a popular item,
and ultimately found one that is most aligned with his observations.
Overall, we found that \toolname{} leads to more informed decision-making as users forage more, as well as more diverse information when considering trade-offs.

\aptLtoX[graphic=no,type=html]{\begin{table}
\small
\centering
\begin{tabular}{p{0.02\linewidth}p{0.4\linewidth}|p{0.45\linewidth}}
\toprule
& \textbf{Statement} & \textbf{Distribution} \\
\midrule
Q1 & Easy to accomplish the task with \toolname{} & \includegraphics{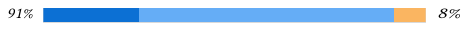}\\
    & Easy to accomplish the task with notebooks & \includegraphics{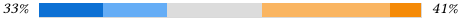}\\
Q2 & Easy to understand trade-offs with \toolname{} & \includegraphics{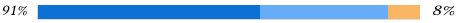}\\
 & Easy to understand trade-offs with notebooks & \includegraphics{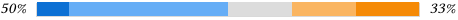}\\
Q3 & Easy to use \toolname{} & \includegraphics{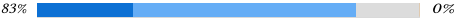}\\
 & Easy to use notebooks & \includegraphics{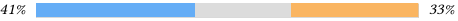}\\
Q4 & Enjoyable to use \toolname{} & \includegraphics{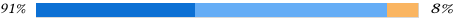}\\
 & Enjoyable to use notebooks & \includegraphics{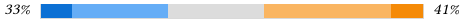}\\
Q5 & Mentally demanding to use \toolname{} & \includegraphics{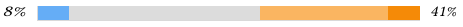}\\
 & Mentally demanding to use notebooks & \includegraphics{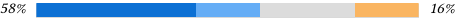}\\
Q6 & Will use \toolname{} in the future & \includegraphics{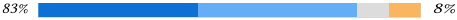}\\
 & Will use notebooks in the future & \includegraphics{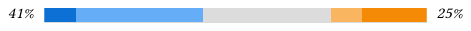}\\
\multicolumn{3}{c}{\includegraphics{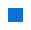} Strongly agree\quad \includegraphics{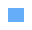} Agree\quad  \includegraphics{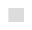} Neutral\quad \includegraphics{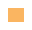} Disagree\quad \includegraphics{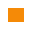} Strongly disagree} \\
\bottomrule
\end{tabular}
\caption{Participants found \toolname{} help them accomplish the task better (91\% vs. 33\%), understand trade-offs easier (91\% vs. 50\%), is easier to use (83\% vs. 41\%), more enjoyable to use (91\% vs. 33\%), and less mentally demanding (8\% vs. 58\%). There are 83\% participants who want to use \toolname{} in the future for similar tasks (vs. only 41\% for notebooks).}
\label{tab:ratings}
\end{table}}{\begin{table*}[t]
\small
\centering
\begin{tabular}{p{0.02\linewidth}p{0.4\linewidth}|p{0.45\linewidth}}
\toprule
& \textbf{Statement} & \textbf{Distribution} \\
\midrule
Q1 & Easy to accomplish the task with \toolname{} & \importancebarchart{0.250}{0.667}{0.000}{0.083}{0.000}{0.250}{91\%}{8\%}\\
    & Easy to accomplish the task with notebooks & \importancebarchart{0.167}{0.167}{0.250}{0.333}{0.083}{0.167}{33\%}{41\%}\\
Q2 & Easy to understand trade-offs with \toolname{} & \importancebarchart{0.583}{0.333}{0.000}{0.083}{0.000}{0.583}{91\%}{8\%}\\
 & Easy to understand trade-offs with notebooks & \importancebarchart{0.083}{0.417}{0.167}{0.167}{0.167}{0.083}{50\%}{33\%}\\
Q3 & Easy to use \toolname{} & \importancebarchart{0.250}{0.583}{0.167}{0.000}{0.000}{0.250}{83\%}{0\%}\\
 & Easy to use notebooks & \importancebarchart{0.000}{0.417}{0.250}{0.333}{0.000}{0.000}{41\%}{33\%}\\
Q4 & Enjoyable to use \toolname{} & \importancebarchart{0.417}{0.500}{0.000}{0.083}{0.000}{0.417}{91\%}{8\%}\\
 & Enjoyable to use notebooks & \importancebarchart{0.083}{0.250}{0.250}{0.333}{0.083}{0.083}{33\%}{41\%}\\
Q5 & Mentally demanding to use \toolname{} & \importancebarchart{0.000}{0.083}{0.500}{0.333}{0.083}{0.000}{8\%}{41\%} \\
 & Mentally demanding to use notebooks & \importancebarchart{0.417}{0.167}{0.250}{0.167}{0.000}{0.417}{58\%}{16\%}\\
Q6 & Will use \toolname{} in the future & \importancebarchart{0.417}{0.417}{0.083}{0.083}{0.000}{0.417}{83\%}{8\%}\\
 & Will use notebooks in the future & \importancebarchart{0.083}{0.333}{0.333}{0.083}{0.167}{0.083}{41\%}{25\%}\\
\multicolumn{3}{c}{\mylegend{Strongly agree}{blue2} \mylegend{Agree}{blue1} \mylegend{Neutral}{gray1}\mylegend{Disagree}{orange1} \mylegend{Strongly disagree}{orange2}} \\
\bottomrule
\end{tabular}
\caption{Participants found \toolname{} help them accomplish the task better (91\% vs. 33\%), understand trade-offs easier (91\% vs. 50\%), is easier to use (83\% vs. 41\%), more enjoyable to use (91\% vs. 33\%), and less mentally demanding (8\% vs. 58\%). There are 83\% participants who want to use \toolname{} in the future for similar tasks (vs. only 41\% for notebooks).}
\label{tab:ratings}
\end{table*}}

\subsection{Finding: \toolname{} Fosters More Thorough Thinking over Trade-offs (RQ3)}

Finally, we found that \toolname{} leads participants to more easily understand trade-offs (91\% vs. 50\%, Q2, Table~\ref{tab:ratings}) and think more about them when they explore the design space.
Participants in the treatment condition mentioned 28 times of trade-offs in total (vs. 15 times in control) when they explored think-aloud.
This is expected, as trade-offs are particularly highlighted in \toolname{} through metrics, hinting users that any design changes are not single-dimensional optimization efforts, but rather require careful balance.
This is well-demonstrated in P11's thought process: \participantQuote{So even for queries with quantities. We've lost something on exact... given how much purchase\_ndcg this cost me on queries with quantities, that doesn't sound super appealing to me}{11}.

\looseness=-1
Such trade-off thinking persists from participants' design exploration to communication:
We found that participants are 28.6\% more likely to explicitly \textit{mention} and \textit{explain} the trade-offs to other stakeholders in their responses with \toolname{},
as exemplified in P5's response to the first task: \participantQuote{There exists trade-off between objectives such as popularity and exact... the weights/objectives suits for keywords with quantities may not perform good on the overall instances}{5}.

\section{Discussion}

\subsection{(Co-)designing Models with Objectives}
In the user study, participants found objectives an important construct for them to navigate through design space.
\toolname{} supports such objective-guided design space navigation very well, with objectives explicitly surfaced and tied to metrics and item rankings.
This, as pointed out, helps participants not only explore design space more efficiently but also explore bigger changes and define more complex objectives.

We found this observation particularly interesting, since users have the same amount of, if not more, freedom to manipulate the objective space in notebooks.
This is only possible because \toolname{} provides an \participantQuote{easy-to-use interface for customized objectives}{9} and more fundamentally, \participantQuote{help defines "playing blocks" in the problem}{4}.
We hypothesize that because \toolname{} surfaces objectives as a first-class citizen, 
it prompts users to think and explore in terms of objectives and effectively encourages them to engage with the objective space beyond small changes.

We believe such an objective-centered design approach can generalize beyond ranker or recommender systems, to any ML models considering multiple objectives.
Recent research on multi-objective fine-tuning~\citep[e.g.,][]{zhou2024beyond,rame2024rewarded} is a good example:
To properly optimize LLMs for multiple objectives, such as safety, coherence, and verbosity, at the same time,
it is also important to understand what objectives to consider and how to trade off objectives when there are conflicts.
Regardless of the exact scenario, \toolname{} serves as a foundation for stakeholders to actively engage in objective design, and comprehensively evaluate different design decisions.

\begin{figure}[t]
    \centering
    \includegraphics[width=0.9\linewidth]{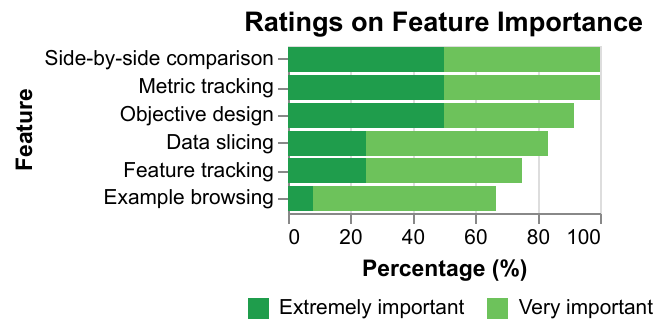}
    \caption{Participants found side-by-side comparison and metric tracking the most important features of \toolname{}, followed by objective design and data slicing.
    }
    \label{fig:importance}
\end{figure}

\paragraph{Co-designing models across stakeholders}
As discussed in Section~\ref{sec:motivation-collab}, model design is also a collaborative effort requiring the participation of different stakeholders.
Our study focuses on a workflow where other stakeholders provide concrete feedback and observations (possibly using \toolname{}) and technical stakeholders take most responsibility for design and evaluation,
while alternative workflows where different stakeholders collaborate in an more iterative and interactive way are not evaluated.

\looseness=-1
We envision that \toolname{} can be used to support participatory design of ML models~\cite{baumer2017toward},
as it can empower less technical stakeholders to understand model design and conduct ``what-if'' analysis through objectives.
By shifting the focus from anecdotal problems to objectives, \toolname{} encourages different stakeholders to talk in the same language and grounds their communication and collaboration in concrete shareable analysis. 
\new{To better support the participatory design settings, future work could further explore how \toolname{} can enable greater agency among less technical stakeholders.
This includes understanding what additional scaffolding or support might be needed to enhance their contribution to the design process and how these contributions could influence model development iteratively.
}

\subsection{Bridging Metrics-centric and Example-centric Mindsets}
In the study, we observed that many participants have a fairly metric-centered mindset -- they agonize over metrics drop and are delighted when metrics increase. 
Such mindsets are prevalent among ML practitioners, but are pointed out to be problematic~\cite{thomas2020problem}.
In the case of ranking or recommendation systems, it is always important to see what users \textit{see} concretely, beyond aggregated metrics improvement.
At the other extreme, less technical stakeholders are often too example-focused, without understanding how the anecdotes can generalize and the greater impact of a fix.

\toolname{} encourages users from two extremes to take a more holistic view, by supporting users to forage comprehensive information, including both metrics and examples, in a unified framework.
For users with example-centric mindsets, \toolname{} always prompts them to think about the larger picture, with metrics and slices information readily available.
For users with metric-centric mindsets, \toolname{} encourages them to look at concrete observations.
Furthermore, \toolname{} can be particularly effective in pushing these users to explicitly rethink about metrics (cf. Section~\ref{sec:rq2}).
Instead of thinking about metrics as something existing and inherently valid, users are encouraged to inspect metrics and design new metrics that \textit{align better with their expectations}, all grounded in concrete observations.

Problems with metrics-centric and example-centric mindsets are repeatedly discussed in the literature:
End users are often found example-focused and make local decisions that can hurt general performance~\cite{wu2019local} -- this is also observed as a major challenge for prompt engineering~\cite{zamfirescu2023johnny}.
Meanwhile, ML experts are found to focus too much on metrics that do not necessarily align with user-facing performance~\cite{gordon2021disagreement}.
The ideas in \toolname{}, from interactive metric design to support for comprehensive information foraging,
can be readily applied to address similar problems in other machine learning scenarios.
We acknowledge that, however, not every scenario would be as straightforward as ranking to gather and present evaluation information to users:
Offline metrics can be hard to compute or unreliable for some problems, 
and more scaffolding might be needed to help users make sense of example results if the model outputs are hard to parse or glance over.
Future work extending \toolname{} might have to identify the best way to help users forage useful information for their evaluation, depending on their scenarios.

\section{Related Work}

\subsection{Multi-objective Machine Learning}
\label{sec:moml}

Multi-objective optimization has gained significant attention in machine learning, especially recommendation and ranking systems, due to the need to balance competing objectives such as relevance, diversity, fairness, and user satisfaction. 
Many techniques have been proposed to address this challenge. 
One approach is label aggregation, which combines multiple labels into a single supervision target for training~\cite[e.g.,][]{dai2011multi,carmel2020multi}. 
Another widely used method is loss aggregation, where different loss functions corresponding to various objectives are merged~\citep[e.g.,][]{hu2018collaborative, mahapatra2023multi,mahapatra2023querywise,tang2024multi}. 
In addition, post-training score aggregation is commonly employed, where outputs from different tasks in multi-task learning frameworks are combined to generate a final ranking that effectively balances these competing goals~\cite[e.g.,][]{zhao2019recommending}. 
Regardless of the exact training method, ML engineers always need to specify appropriate objective formulations and their weights, in order to control the priorities of different objectives to design the final ranking experience, 
and \toolname{} can serve as a framework for ML engineers to explore and evaluate different ranker designs.

Beyond recommendation and ranking systems, many machine learning problems are also effectively multi-objective,
as researchers have increasingly paid attention to model qualities beyond accuracy~\cite{ribeiro-etal-2020-beyond}: from fairness~\cite{shah-etal-2020-predictive}, robustness~\cite{goel-etal-2021-robustness}, to safety~\cite{DeepRoad}.
However, most research here focuses on one single (additional) objective at a time, and 
commonly relies on additional data augmentation~\cite[e.g.,][]{nl-augmenter} or more comprehensive data curation (e.g., for LLM instruction-tuning~\cite{ouyang2022training}). 
More recently, multi-objective fine-tuning~\citep[e.g.,][]{zhou2024beyond,rame2024rewarded} has been proposed to optimize for multiple objectives, such as safety, coherence, and verbosity, at the same time, for LLM generation.
We envision that \toolname{} can be used as a foundational framework for trading off objectives for general multi-objective machine learning problems.

\subsection{Pitfalls of Recommender Systems}
Recommender systems are known to exist a series of different biases~\cite{chen2023bias}:
There is selection bias~\cite{marlin2012collaborative}, where user behavioral signals are sparse and often missing non-randomly, leading to biased predictions.
There is position bias~\cite{collins2018study}, where users tend to interact with top items in the list --
this can cause a self-reinforcing feedback loop where top items stay at the top with more user interactions.
There is also popularity bias~\cite{klimashevskaia2024survey}, where popular items are recommended even more frequently than their popularity -- 
this can reduce the visibility of other items and hurt fairness.
Optimizing solely for user behavioral signals can also cause undesired outcomes like filter bubbles~\cite{pariser2011filter} or encouraging radicalization~\cite{ribeiro2020auditing}.

While there are many research works on debiasing recommenders and counterfactual learning~\cite[e.g.,][]{joachims2017unbiased,agarwal2019general,ai2018unbiased,xiao2023towards,luo2023model}, few have demonstrated the benefits in production environments~\cite[e.g.,][]{zou2022large,gupta2024unbiased}. 
Nowadays, it is common to inject extra prior knowledge, such as relevance, fairness, and diversity, into the objectives to mitigate biases~\cite[e.g.,][]{wu2022multi, zheng2022survey,momma2020multi,mahapatra2023querywise}.
\toolname{} can be used by practitioners to identify what objectives can alleviate the biases, and also decide the trade-off between additional objectives and main objectives.

\subsection{Interactive Systems for Machine Learning}
Interactive machine learning~\cite{fails2003interactive} aims to include humans in ML model construction procedures, by helping humans understand model failures and suggest improvements~\cite[e.g.,][]{amershi2015modeltracker, krause2014infuse}.
Researchers have explored different forms of human feedback, from labeling~\cite{heimerl2012visual} to feature selection~\cite{krause2014infuse} for model improvement, which has been found to cause local decision pitfall~\cite{wu2019local},
where users over-generalize from a single observation.
Instead of helping non-experts create better models with one single objective as in interactive machine learning,
\toolname{} enables stakeholders to understand and explore trade-offs among multiple objectives.
\toolname{} also avoids local decision pitfalls by presenting users with diverse evaluation information at both the example level and aggregation level.

Another closely related area is tooling support for machine learning.
For the evaluation side, existing work has explored designing tools and interfaces to support error analysis~\cite{wu2019errudite, wu2020tempura}, data slicing~\cite{cabrera2023zeno, suresh2023kaleidoscope}, model testing~\cite{ribeiro-etal-2020-beyond, ribeiro2022adaptive, yang2023beyond}, as well as LLM-powered evaluation~\cite{kim2024evallm}.
For the design side, model sketching~\cite{lam2023model} enables practitioners to author models from high-level concepts.
ConstitutionMaker~\cite{petridis2024constitutionmaker} supports users to convert their feedback to constitutions for chatbots.
There is also work on LLM chaining~\cite{wu2022ai, arawjo2024chainforge} to support users to design LLM workflows.
\toolname{} borrows ideas from existing work on evaluation (e.g., data slicing), but focuses on the problem of effectively presenting comprehensive evaluation information to users.
Compared to existing work for model design, \toolname{} is the first to target multi-objective problems and explicitly surface objectives as the key design construct.

\section{Conclusion}
In this work, we present \toolname{}, a framework that places objectives at the center of the model design process. 
\toolname{} allows users to directly engage with objective space, enabling real-time exploration and evaluation of design trade-offs.
Our evaluation shows that \toolname{} helps practitioners explore the design space more efficiently, make more informed decisions, and are more aware of the inherent trade-offs.
\toolname{} opens new avenues for an objective-centric design process applicable to other multi-objective machine learning problems, 
as well as sheds light on future designs that encourage practitioners to think beyond only metrics or examples for evaluation.

\begin{acks}
We thank Ram Kandasamy, Subhajit Sanyal, Vivek Mittal, Behzad Tabibian, Dhivya Eswaran, Gowri Raman, Anshuka Rangi, Holakou Rahmanian, Hancao Li, Qing Jin, Yinuo Ren, Xun Tang, and others for their feedback on this work. 
\end{acks}

\bibliographystyle{ACM-Reference-Format}
\bibliography{main}

\appendix

\section{Post-study Survey Questions}
\label{sec:survey}

\begin{itemize}
    \item Please rate the following in terms of how much you agree or disagree with each statement. (Statements listed in Table~\ref{tab:ratings})
    \item How important were these aspects of working with \toolname{}? (Aspects listed in Figure~\ref{fig:importance})
    \item For similar tasks in the future, which tool do you prefer using? Why?
    \item What stood out to you about the experience of using \toolname{}? For example, was anything good, bad, surprising, or notable?
    \item If you want to use \toolname{} in the future, what is a scenario you want to use it for?
    \item If there is one thing you can change about \toolname{}, what would you change? (Feel free to write more if you want)
\end{itemize}

\end{document}